\documentclass[preprint2]{aastex631}
\usepackage{amsmath,amsfonts,amssymb}
\usepackage{bm}
\usepackage{xspace}
\usepackage{listings}

\graphicspath{{Figures/}}

\newcommand{\IRSSquare}{IRS$^2$\xspace}
\newcommand{\electrons}[1]{#1~$e^-$\xspace}

\begin{document}

\title{NSClean: An Algorithm for\\ Removing Correlated Noise from JWST NIRSpec Images}

\correspondingauthor{Bernard J. Rauscher}
\email{Bernard.J.Rauscher@nasa.gov}

\author[0000-0003-2662-6821]{Bernard J. Rauscher}
\affiliation{NASA Goddard Space Flight Center\\
Observational Cosmology Laboratory\\
8800 Greenbelt Road\\
Greenbelt MD 20771, USA}

\begin{abstract}
NSClean is an algorithm and python package for removing faint vertical banding and ``picture frame noise'' from JWST Near Infrared Spectrograph (NIRSpec) images. NSClean uses known dark areas to fit a background model to each exposure in Fourier space. When the model is subtracted, it removes nearly all correlated noise. Compared to simpler strategies like subtracting the rolling median, NSClean is more thorough and uniform. NSClean is computationally undemanding, requiring only a few seconds to clean an image on a typical laptop. The NSClean package is freely available from the NASA JWST website \citep{jwst2023}.
\end{abstract}

\section{Introduction} \label{sec:intro}

JWST is today's premier space observatory for mid and near-infrared (NIR) astronomy \citep{Gardner2023}. To enable science objectives cutting across astrophysics, JWST carries a suite of four science instruments:  a Near Infrared Camera \citep[NIRCam;][]{Rieke2023}, a Near Infrared Imager and Slitless Spectrograph \citep[NIRISS;][]{Doyon2023}, a Mid-infrared Instrument \citep[MIRI;][]{Wright2023}, and a Near Infrared Spectrograph \citep[NIRSpec;][]{Jakobsen_2022}. This article concerns NIRSpec, an algorithm, and a software package to further reduce its already low read noise: ``NSClean''.

From early on, it was understood that NIRSpec required ultra-low noise detectors. NIRSpec is detector noise limited for all but prism-mode observations. This is in contrast to other JWST instruments that are generally limited by the astronomical background. Consequently, NIRSpec had lower noise requirements than other JWST instruments.

``Total Noise'' is a concept that was introduced for JWST. It is intended to account for the combined effects of detector read noise and shot noise on integrated dark current. To measure it; one defines a standard scientific exposure, takes many such exposures (typically $>$40), and then computes the standard deviation per pixel. Across JWST's NIR instruments, the exposure time was taken to be 1000 seconds. For NIRCam and NIRISS, median total noise was required to be $<$\electrons{10}  per exposure. For NIRSpec, the requirement was $<$\electrons{6}.\footnote{MIRI uses a different detector technology for which the comparison is not relevant.}

This $<$\electrons{6} noise requirement is the reason why we developed Improved Reference Sampling and Subtraction \citep[\IRSSquare; pronounced IRS-square;][]{Rauscher_2017}. In \IRSSquare mode, NIRSpec uses a special clocking pattern and reference correction pipeline step to reduce correlated noise as far as possible using the NIRSpec detector's built-in references. Using \IRSSquare, NIRSpec's total noise is slightly less than \electrons{6} on average, and to within the uncertainties compliant with requirements. \IRSSquare is the recommended readout mode for most observations except for extremely bright targets \citep{2016jdox}.

However, even with NIRspec's detectors meeting requirements, many NIRSpec observers report seeing faint, correlated read noise in count rate images that complicates calibration. Fortunately, for NIRSpec, much of this can be removed using dark areas of images as references.

\begin{figure*}[t]
\centering
\includegraphics[width=\textwidth]{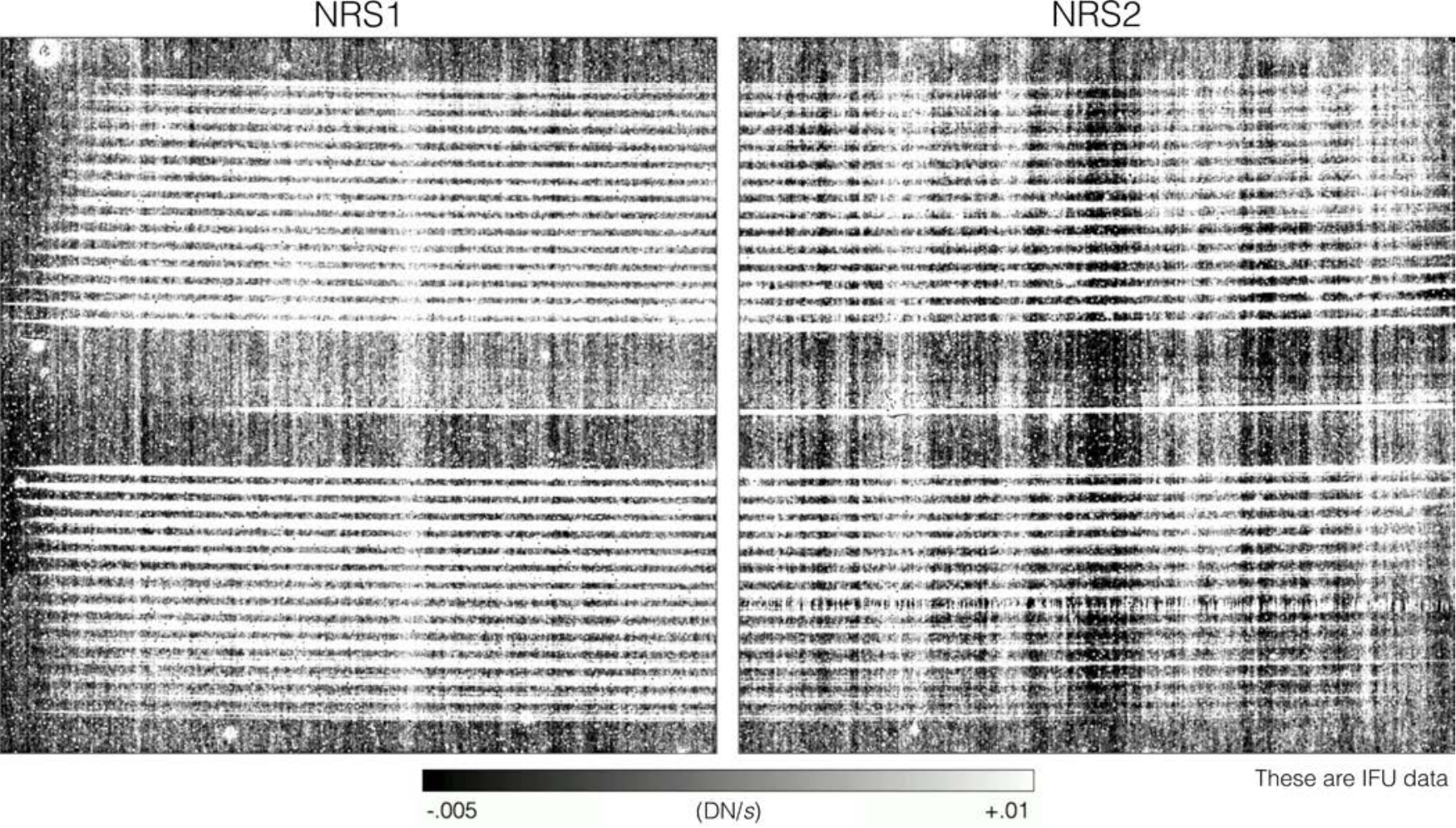}
\caption{The JWST pipeline makes NIRSpec count rate images like those shown here. This observation used NIRSpec's IFU mode which produces 30 horizontal spectral traces per detector. To highlight correlated noise, we have smoothed the images and set the greyscale roughly equal to NIRSpec's 6~electrons total noise requirement. One sees vertical banding in the central regions of both detectors. Toward the edges of both detectors, there seems to be less correlated noise. This is the ``picture frame''. While both types of residual noise are less than NIRSpec's total noise requirement, they nevertheless complicate calibration. For example, they can produce negative fluxes and features that mimic emission lines or continuum. NSClean fits a background model to dark areas of each exposure and subtracts it to remove picture frame noise and vertical banding.}\label{fig:nrs1_and_nrs2_pipeline}
\label{default}
\end{figure*}

Figure~\ref{fig:nrs1_and_nrs2_pipeline} shows an example of the correlated noise from an early NIRSpec Integral Field Unit (IFU) observation. We have smoothed the images and stretched the greyscales to emphasize correlated noise that would otherwise be more difficult to see against the background of NIRSpec's $\sim$6 electrons total noise. One sees a ``picture frame'' effect, whereby areas near the edges of both detectors on all four sides  seem less noisy. In the interiors, one sees faint vertical striping. While the amplitude is small, this correlated noise can undermine accurate photometry when no local sky is available and cause false spectral features when working near the read noise.

\begin{figure*}[t]
\centering
\includegraphics[width=\textwidth]{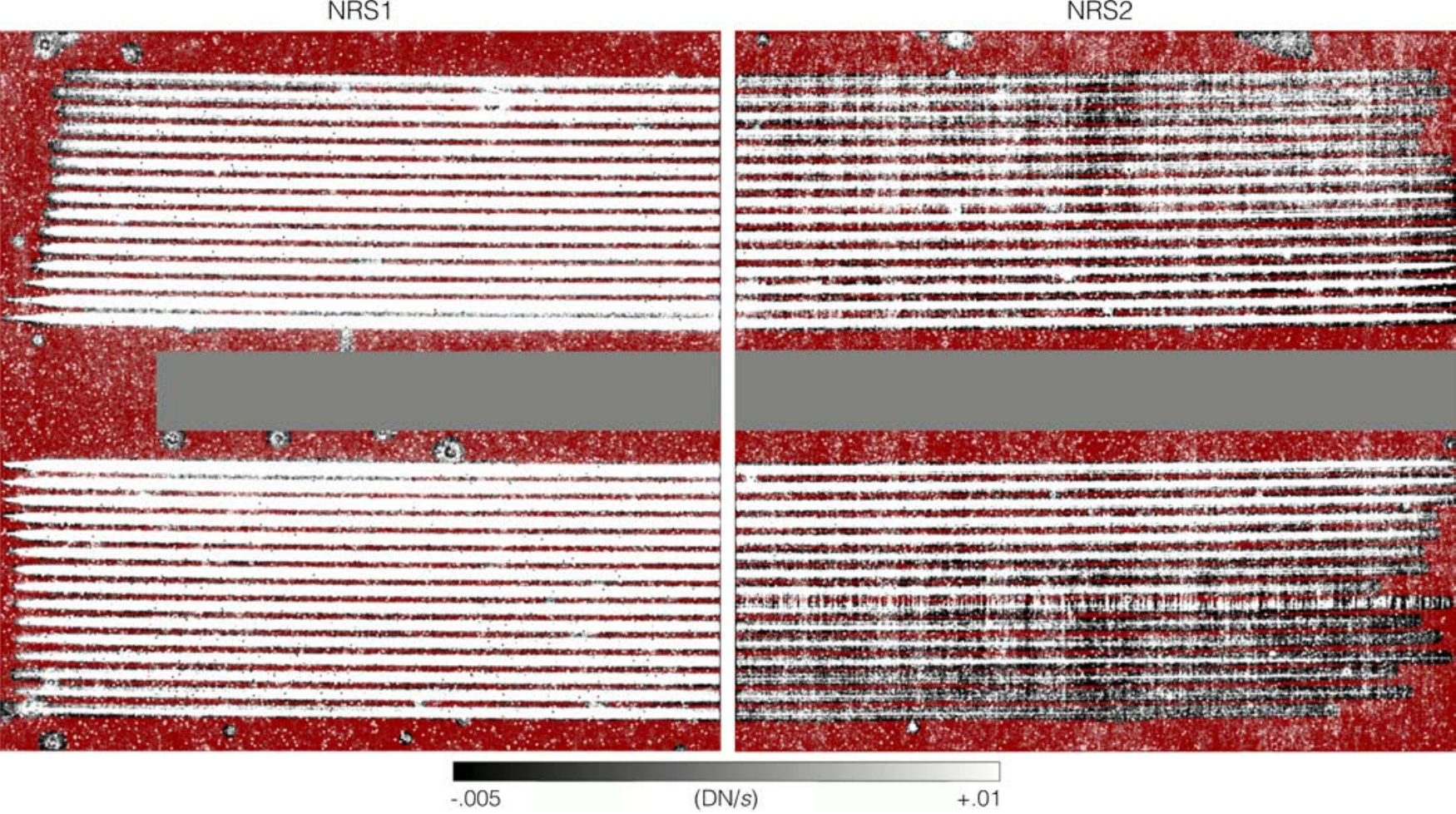}
\caption{We used the background masks shown here for development. The underlying grayscale image is the median of a stack of illuminated IFU exposures. The 30 spectral traces per detector are clearly visible. We used the red-shaded pixels to make the background model. As described in the text, we used the GNU Image Manipulation Program (GIMP) to manually make the masks since we only needed one set to write the software. Section~\ref{sec:making-masks-templates} describes how mask making can be automated. The grey rectangles blank off potentially illuminated (by scattered light) areas of the focal plane that we left unconstrained during background fitting.}\label{fig:mask}
\label{default}
\end{figure*}

\begin{figure*}[t]
\centering
\includegraphics[width=\textwidth]{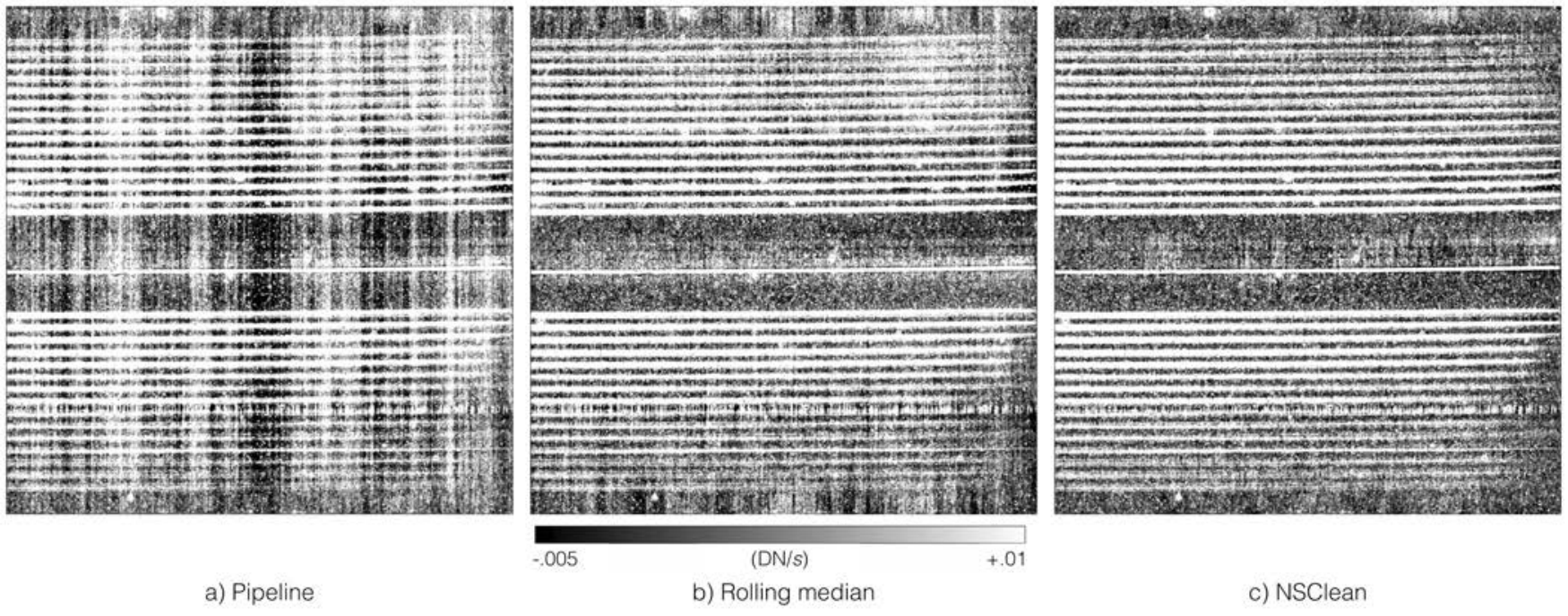}
\caption{This figure shows the a) correlated noise that is visible in pipeline calibrated images. The actual pipeline products do not look this bad. We have adjusted the grayscale and blurred the images slightly to highlight correlated read noise. Panel b) shows the effect of subtracting the median of a few neighboring columns from each column. Finally, panel c) shows the NSClean result. Panels b and c are noticeably cleaner than panel a. Comparing panels b and c, panel c shows more uniform and complete background subtraction. }\label{fig:results}
\label{default}
\end{figure*}

NSClean uses blanked off areas of NIRSpec scenes to model and subtract the background, including the correlated noise. Figure~\ref{fig:mask} shows a typical background pixels mask for IFU mode. The red shaded pixels are used to build the background model. Because it uses more information and allows for structure in the background, NSClean's correlated noise correction is more complete and more uniform than is possible using simpler techniques such as subtracting a rolling median of background columns (Figure~\ref{fig:results}).

We developed and tested NSClean using IFU data. As Figure~\ref{fig:results} shows, the reduction in correlated noise can be dramatic. Although we have not tested NSClean with multi-object spectrograph (MOS) data ourselves, we understand from colleagues on the Early Release Science (ERS) TEMPLATES\footnote{TEMPLATES stands for Targeting Extremely Magnified Panchromatic Lensed Arcs and their Extended Star formation. The TEMPLATES team is co-led by J. Rigby of NASA Goddard Space Flight Center and J. Vieira of the University of Illinois at Urbana-Champaign.} team that it works well. They are studying  extremely magnified panchromatic lensed arcs with extended star formation. The advantage for TEMPLATES is that the irregularly shaped sources seldom align with a standard 3-slit MOS dither pattern. We are in the early stages of testing a Bright Object Time Series (BOTS) module now. We plan to include that in a future release.

The rest of this paper is structured as follows. In Section~\ref{sec:cause}, we explain the underlying physical cause of the correlated noise that NSClean removes. Section~\ref{sec:algorithm} describes the NSClean algorithm. The idea is simple. NSClean approximates the Fourier transform of the background using an algorithm that is robust against missing data and gaps where spectra lie. NSClean then anodizes the Fourier transform using a low-pass filter\footnote{The apodizer has unity gain at low frequencies. After about 9 frequencies, its gain is rolled to zero using one half of a cosine wave.} and inverts it to compute the background model. Section~\ref{sec:implementation} discusses the python-3 implementation and computing requirements. We close in Section~\ref{sec:summary} with a summary.

\section{Physical Cause of the Correlated Noise}\label{sec:cause}

Our focus in this paper is on the specific correlated noise that NSClean is designed to fix. Readers who want to learn more about NIRSpec's read noise in general may want to see some of our earlier papers. \citet{Rauscher_2015} describes the origins of NIRSpec's white and $1/f$ noise, and provides a python package for simulating it. \citet{Rauscher_2017} describes NIRSpec's \IRSSquare readout mode. Without \IRSSquare, the residual correlated noise that remains today would be much worse.

The correlated noise that remains after \IRSSquare  is a logical consequence of how \IRSSquare works. NIRSpec uses two Teledyne H2RG NIR detector arrays \citep{Loose:2003vh}. Each H2RG provides two types of reference information that can be used to remove correlated read noise. These are the ``reference pixels'' that form a 4-pixel wide frame on all sides of NIRSpec images and one ``reference output'' per H2RG. The reference output is not visible in the usual  pipeline data products, but it is used most of the time. As described in \citet{Rauscher_2017}, \IRSSquare is built on principal component analysis (PCA) showing that NIRSpec's read noise is covariance stationary to a high degree of approximation. Informally, this means that the read noise is independent of when one looks.

It turns out that in JWST's NIR detector systems, thermal instability causes noise that is not covariance stationary. There is a picture frame pattern that changes in time at the $\sim$\electrons{1} level. \citet{Rauscher_2013} describe how small temperature fluctuations can drive the picture frame. This is why the vertical banding that is visible in Figure~\ref{fig:results}a seems to fade away near the edges. The relatively quiet edges are in the picture frame while the vertical bands are not. \IRSSquare relies on the reference pixels to see noise in order to remove it. Since the reference pixels are in the picture frame and do not see the vertical banding, \IRSSquare is powerless to remove it. 

\section{Algorithm}\label{sec:algorithm}

NSClean is built on the Fourier transform of the instrumental background. Our treatment starts in Section~\ref{sec:fft}, by reviewing how python's numpy package implements the classical Fast Fourier Transform \citep[FFT;][]{Cooley1965} for fully sampled data. Since NIRSpec's background is not fully sampled (because of astronomical sources), Section~\ref{sec:nsclean-ft} explains how NSClean computes a statistically optimal approximation to the Fourier transform using all available background samples.

The next two subsections describe the linear algebra that underpins NSClean. Insofar as possible, we have tried to use a consistent, standard notation. Throughout this paper, boldface lowercase letters are vectors and uppercase boldface letters are matrices. When discussing matrix elements, we use superscripts for row indices and subscripts for column indices.

\subsection{Numpy's Classical FFT}\label{sec:fft}

For dark exposures, one can use numpy's FFT package to compute the Fourier transform of an image column. Like all FFTs, numpy uses a highly-efficient factorization of the Fourier matrix, $\mathbf{F}$, to solve the matrix equation,
\begin{equation}
\mathbf{F} \mathbf{f} = \mathbf{d},\label{eq:ft1}
\end{equation}
where $\mathbf{f}$ is the Fourier transform of the data, $\mathbf{d}$. For $n$ pixels per column, in numpy the elements of $\mathbf{F}$ are,
\begin{equation}
F^m_k = \exp\left\{2\pi i\frac{m k}{n}\right\}.
\end{equation}
Because NIRSpec's data are real valued and $n=2048$ is an even number; $m=0,1,\dots, n-1$ and $k=0,1,\dots,n/2$.

\subsection{NSClean's Fourier Transform}\label{sec:nsclean-ft}

For NIRSpec's incompletely sampled background, NSClean uses weighted least squares to approximate Fourier transforms. The starting point is again equation~\ref{eq:ft1},
\begin{equation}
\mathbf{F} \mathbf{f} \approx \mathbf{d},\label{eq:ft2}
\end{equation}
 but now as an approximation and with the understanding that $\mathbf{F}$, $\mathbf{f}$, and $\mathbf{d}$ are incomplete. $\mathbf{F}$ is missing columns where light falls on the detector and rows for frequencies that we choose not to fit. $\mathbf{f}$ contains only a few very low frequencies to minimize noise. $\mathbf{d}$ is missing rows where the detector is illuminated.
 
 To solve equation~\ref{eq:ft2} using least squares, we minimize the generalized distance squared,
 \begin{equation}
\delta^2 = (\mathbf{F} \mathbf{f} -\mathbf{d})^{\rm H} \mathbf{W}(\mathbf{F} \mathbf{f} -\mathbf{d}),\label{eq:distance}
\end{equation}
using all available background samples. The symbol, ``$^{\rm H}$'', denotes the conjugate transpose, which is also known as the Hermitian transpose. A weight matrix, $\mathbf{W}$, is required to compensate for non-uniform background sampling. NSClean weights inversely by the local sample density squared, $\rho^{-2}$:
\begin{equation}
\mathbf{W}=\begin{bmatrix}
\rho_{0 0}^{-2} & 0 & 0& 0\\
0 & \rho_{1 1}^{-2} & 0& 0\\
0 & 0& \ddots & 0\\
0& 0& 0& \rho_{n^\prime-1\hspace{6pt} n^\prime-1}^{-2}
\end{bmatrix}.
\end{equation}
$\mathbf{W}$ is diagonal and equal to its conjugate transpose. Section~\ref{sec:W} describes $\mathbf{W}$ in more detail. The quantity $n^\prime \leq n$ is equal to the number of background samples. Under these conditions, the least squares solution to Equation~\ref{eq:distance} is,
\begin{equation}
\mathbf{f}=\left(\mathbf{W}^{1/2}\mathbf{F}\right)^+\mathbf{W}^{1/2}\mathbf{d}.\label{eq:soln}
\end{equation}
The symbol, ``$^+$'', denotes the Moore-Penrose inverse. Being a Fourier transform, the quantity $\mathbf{f}$ is a complex valued vector.

Equation~\ref{eq:soln} is this paper's key result. NSClean uses this expression to approximate the Fourier transform of the incompletely sampled background.

\begin{figure*}[t]
\centering
\includegraphics[width=\textwidth]{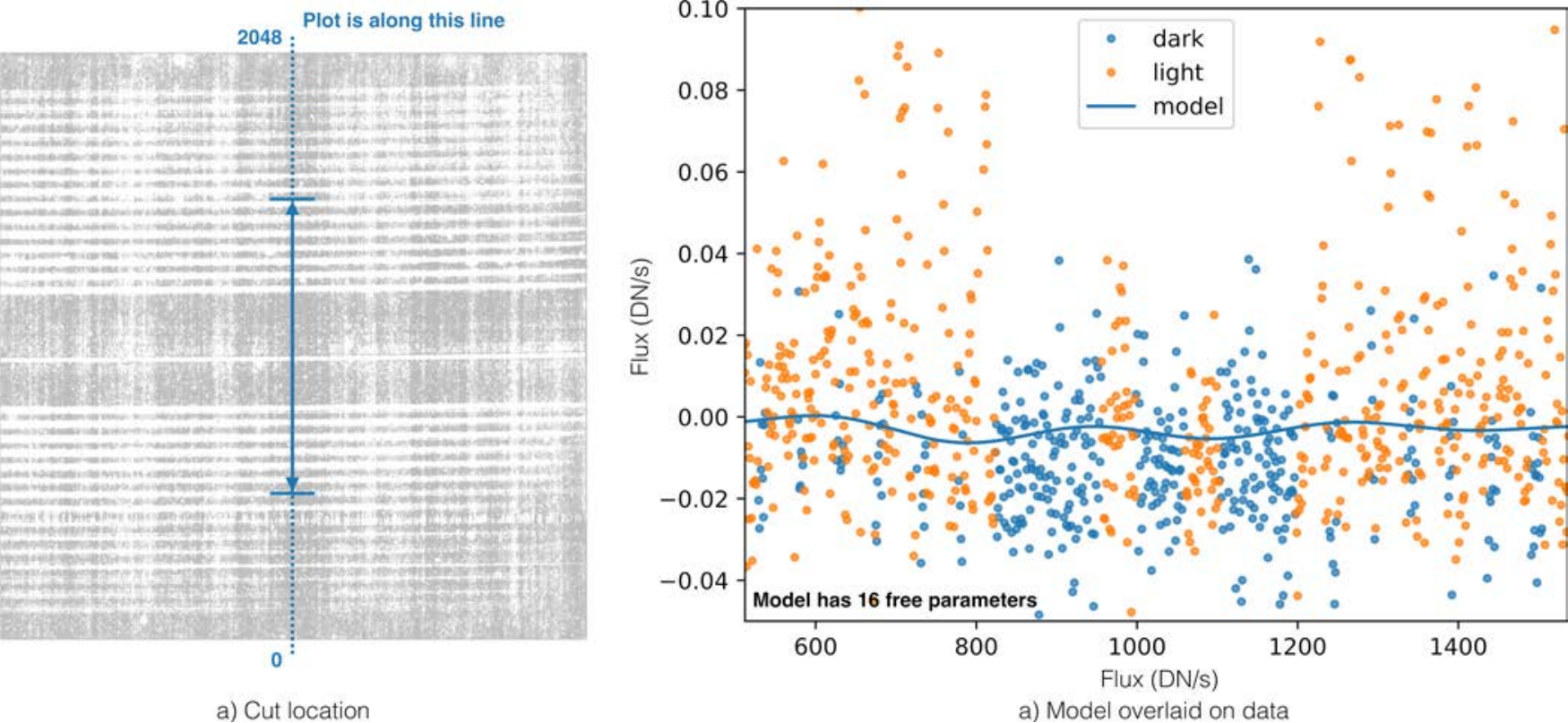}
\caption{This figure shows the background model for one image column. Panel a) shows a vertical cut through NRS2. This channel has the most correlated noise of the two detectors. To show pixel-level detail, Panel b) shows only the innermost 1024 rows. The blue points are background samples and the orange points are pixels that the background mask marked as potentially illuminated. The blue line is the background model. It correctly passes near the centers of groups of blue points. The model is smooth, continuous, and very low noise compared to the pixels themselves.}\label{fig:example}
\label{default}
\end{figure*}

Figure~\ref{fig:example} shows an example of how equation~\ref{eq:soln} works in practice. Panel a) shows a vertical cut through NRS2, which has the most correlated noise of the two detectors. To show detail, Panel b) plots only the innermost 1024 rows. The blue points are background samples, the orange points are pixels that the background mask marked as potentially illuminated, and the blue line is the model built using equation~\ref{eq:soln}. As a practical matter, we were able to fit about nine frequencies ($\approx 18$ free parameters) before we started to see increased noise due to over fitting. As expected, the blue line passes near the centers of groups of blue points. It is smooth, continuous, and very low noise compared to the pixels themselves.

\subsection{The Weight Matrix, $\mathbf{W}$}\label{sec:W}

The weight matrix compensates for uneven background sampling. Returning to Figure~\ref{fig:mask}, there are often only a few rows of blanked off background pixels between the spectral traces. But; near the bottom, middle, and top of each detector, there are  much larger areas of background pixels. When nothing is done to compensate for the uneven background sampling, scientifically uninteresting areas of the scene carry far too much weight.

As described earlier, NSClean computes the Fourier transforms of columns individually using weighted least squares fits. After a bit of trial and error, we found that weighting inversely by the local background sample density in columns works well. There is nothing fundamental about this weighting scheme. We imagine that some observers may find better ones for their data. 

\begin{figure}[t]
\centering
\includegraphics[width=.47\textwidth]{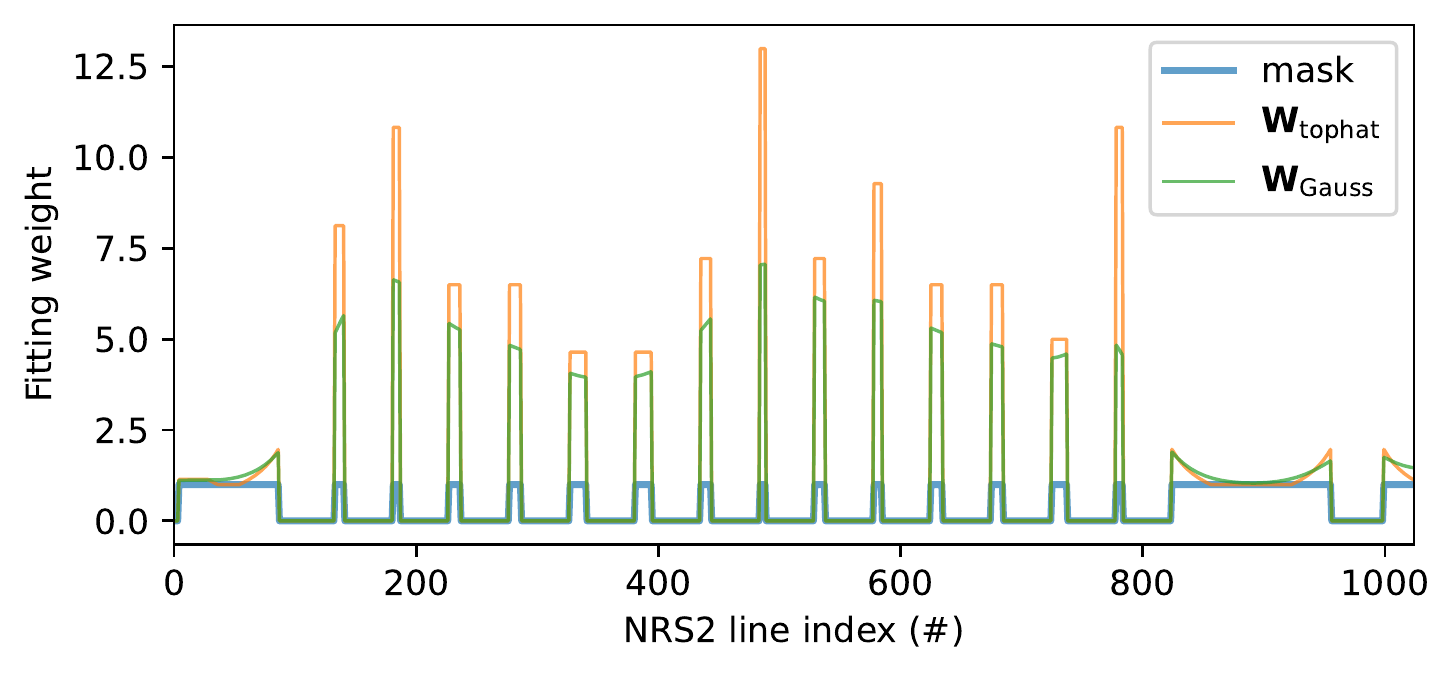}
\caption{This figure shows the background mask and diagonal of $\mathbf{W}^{1/2}$ along the same vertical cut through NRS2 that is shown in Figure~\ref{fig:example}a. For clarity, we show only the first 1024 rows. Mask values =1 are treated as background and mask values =0 are treated as potentially illuminated. The orange curve shows the weights that result from convolving a 65 pixel wide tophat. The green curve shows that weights derived from convolving a Gaussian kernel with $\sigma=32~\textrm{pixels}$. As described in the text, NSClean uses Gauss-convolution because the resulting weights are more uniform and the weight curve is not quantized.}\label{fig:W}
\label{default}
\end{figure}

One could compute the local sample density by convolving the background mask with a tophat function (Figure~\ref{fig:W}). While effective, the resulting weight curve is quantized in units of the tophat's width. To eliminate the quantization while still approximating the local density, NSClean convolves columns of the background mask with a Gaussian kernel. In the current release, the kernel's standard deviation is hard coded to be $\sigma=32~\textrm{pixels}$. Going forward, it may be possible to come up with something more elegant. 32~pixels seems to work well for many IFU observations.

\subsection{Making Masks}\label{sec:making-masks}

This section describes how we made the masks shown in Figure~\ref{fig:mask}. Since we only needed two masks for development purposes, we made them manually. For real science programs, it would be helpful to automate the process. The TEMPLATES  team has tools that create masks from information available in the pipeline. This is described in Section~\ref{sec:making-masks-templates}.

For this paper, we used the GNU Image Manipulation Program (GIMP) to make masks. The starting point was a NIRSpec ``rate.fits'' image displayed in SAOImage DS9. We adjusted the grayscale to show  spectral traces and illuminated areas. Then, using DS9, we exported the 2048$\times$2048 image to Portable Network Graphics (PNG) format. We chose PNG because we knew that it imported well into GIMP.

In GIMP, we created a selection that contained all illuminated pixels and inverted this to get the background sample. We shaded the selection 30\% red as shown in Figure~\ref{fig:mask} and exported the result to another PNG image. The NSClean package includes a python method that is capable of importing a shaded PNG image and converting it to a FITS background pixels mask. There is an example notebook in the distribution that shows how to do this.

\subsection{Automated Mask Making}\label{sec:making-masks-templates}

This description was provided by Taylor Hutchison and Brian Welch of the TEMPLATES team. It has been lightly edited for consistency.

For the TEMPLATES overview paper \citep{Rigby2023}, the team created masks from the pre-existing stage 2 products (specifically the cal.fits files). These mask out everything except the science pixels.\footnote{If the pipeline has not been run before, these stage 2 cal.fits files can be found and downloaded from the \href{https://mast.stsci.edu/portal/Mashup/Clients/Mast/Portal.html}{MAST archive portal}.}  In addition to the masking provided by the cal.fits files, Rigby~\textit{et~al.} masked out the region of the detector where the fixed slits are located and all jump detections (using the data quality flag).  Finally, to ensure clean masking, they employed a subtle binary dilation to the cal.fits files to add a small amount of buffering on the mask edges.

\section{Implementation}\label{sec:implementation}

NSClean is written in python-3. We chose python for compatibility with the rest of the JWST pipeline. The current NSClean version is not computationally demanding. The typical cleaning time for one 2048$\times$2048 NIRSpec image is a few seconds. This assumes that multithreading is turned on for the python linear algebra libraries as described in Section~\ref{sec:multithread}.

The current NSClean version works column-by-column. Since there are only 2048 pixels per column, this means that it requires very little RAM, and the time penalty for projecting out Fourier vectors using Equation~\ref{eq:soln} is small compared to using the FFT algorithm.\footnote{The FFT  only works for fully sampled data, which we do not have.}

\subsection{Computing Requirements}\label{sec:requirements}

The execution time on our development server is about 6~seconds for one 2048$\times$2048 pixel NIRSpec image. The server, which is a few years old, has 8$\times$ Intel Xeon cores running at 3.5~GHz and 250~GB of RAM. In practice, NSClean uses only a tiny fraction of the RAM. Although our server has an NVIDIA Quadro M4000 GPU with 8 GB of RAM, in practice we found that NSClean's execution time was about the same in CPUs as in the GPU. This is because Equation~\ref{eq:soln} 's matrices are not large when images are processed in columns.

We have also tested NSClean on a 2019 MacBook Pro. Execution time on the MacBook is about 12 seconds per image. The MacBook has an 8-Core Intel i9 CPU running at 2.3~GHz and 32 GB of RAM. Again, NSClean did not use much of this RAM. According to the Apple ActivityMonitor App, peak usage was about 150 MB.

Our development server had the following software; Oracle Linux Server release 8.7, python-3.10.8, astropy-5.0.4, cupy-11.5.0, numpy-1.22.3, and pillow-9.3.0.

\subsection{Multithreading}\label{sec:multithread}

NSClean is not explicitly multithreaded. In practice, however, we always have multithreading turned on for python's linear algebra libraries. As a result, when we run NSClean, it usually shows all CPUs being used because most of the work is linear algebra.

On our Intel-based computers, this is done by installing the Intel version of numpy and setting an environment variable. For our 8-core server, the python code is as follows.
\begin{lstlisting}[language=Python,basicstyle=\footnotesize]
  import os
  os.environ[`MKL_NUM_THREADS']=`8'
\end{lstlisting} 
Our understanding is that on non-Intel computers, similar functionality exists, although the environment variables are different.

When a GPU is used, python's cupy package automatically parallelizes  linear algebra over however many GPU cores are available. Our NVIDIA Quadro M4000 has 1664 CUDA cores. However, as a practical matter, on our server the time penalty for uploading/downloading data to the GPU overcame the advantages provided by increased parallelization. We therefore typically run using only the system's CPUs.

\subsection{Installing NSClean}\label{sec:download}

NSClean is a standard pip-installable python package. It is available from the NASA JWST website \citep{jwst2023}. To install it on MacOS or Linux, change into a directory that is in your python path, and download the distribution. Then, use pip to install it,
\begin{lstlisting}
  pip install -e nsclean.
\end{lstlisting}
This will install nsclean as an editable package in your python path.

\section{Summary}\label{sec:summary}

Many JWST observers are finding that there is faint vertical banding and a picture frame pattern in pipeline calibrated NIRSpec images. The effect is particularly challenging for IFU observations because it can add spectral features that are not real. This article describes the NSClean python package that uses dark areas of NIRSpec scenes to remove this noise. To use NSClean, one must provide a mask specifying which pixels are to be treated as background. For each count rate image, NSClean then: (1) computes the Fourier transform of the background using an algorithm that can handle missing data, (2) applies a low-pass apodizing filter to reduce noise, and (3) inverts the Fourier transform yielding a background model. When the background model is subtracted from the image, it removes most of the correlated noise. NSClean is simple and computationally undemanding. The NSClean python package is freely available for download from the NASA JWST Website \citep{jwst2023}.

\begin{acknowledgments}
This work was supported by NASA as part of the JWST Project. I wish to thank Jane Rigby of the TEMPLATES team for her eagerness to test early versions of NSClean and enthusiastic support ever since. Jane read an early draft of this article and made many helpful comments. I am grateful to TEMPLATES team members Taylor Hutchison and Brian Welch for providing the description of how TEMPLATES makes background masks (Section~\ref{sec:making-masks-templates}) on very short notice. I thank Stephan Birkmann of the JWST NIRSpec Team. Stephan was the first person to describe the issue to me and its impact on early JWST science. I have worked with Stephan ever since the early days of NIRSpec. I has always been, and continues to be, a pleasure.
\end{acknowledgments}

\vspace{5mm}
\facilities{JWST(NIRSpec)}

\software{astropy \citep{2013A&A...558A..33A,2018AJ....156..123A}}




\bibliography{pasp_article}{}
\bibliographystyle{aasjournal}

\end{document}